\documentstyle [12pt,twoside]{article}
\newskip\humongous \humongous=0pt plus 1000pt minus 1000pt
\def\caja{\mathsurround=0pt}

\newif\ifdtup
\def\panorama{\global\dtuptrue \openup2\jot \caja
        \everycr{\noalign{\ifdtup \global\dtupfalse
        \vskip-\lineskiplimit \vskip\normallineskiplimit
        \else \penalty\interdisplaylinepenalty \fi}}}

\def\eqalignnotwo#1{\panorama \tabskip=\humongous
        \halign to\displaywidth{\hfil$\displaystyle{##}$
        \tabskip=0pt&$\displaystyle{{}##}$
        \tabskip=\humongous&\llap{$##$}\tabskip=0pt
        \tabskip=0pt&$\displaystyle{{}##}$\hfil
        \crcr#1\crcr}}

\def\begintitle#1#2#3#4
        {\begin{titlepage}
         \centerline{#1 \hfill UMDGR-92-#2}
         \begin{center}\vglue .4in
         {\large\bf #3}\\[.4in]
         {\bf #4}\\[.2in]
         {\it Department of Physics}\\
         {\it University of Maryland, College Park, MD 20742}\\[.4in]
         {\bf ABSTRACT}\\
         \end{center}
         \begin{quotation}}
\def\endtitle
         {\end{quotation}
          \end{titlepage}
          \newpage}

\def\cC{{\cal C}}\def\cD{{\cal D}}
\def\cG{{\cal G}}

\def\cS{{\cal S}}
\def\cV{{\cal V}}

\def\d{\delta}\def\e{\epsilon}
\def\l{\lambda}
\def\u{\mu}\def\r{\rho}\def\s{\sigma}

\def\L{\Lambda}\def\S{\Sigma}

\def\pd{\partial}

\def\otw{\widetilde}
\def\utw#1{\rlap{\lower1ex\hbox{$\sim$}}#1{}}
\def\pb#1{\rlap{\lower1ex\hbox{$\leftarrow$}}#1{}}
\def\pf#1{\rlap{\lower1ex\hbox{$\rightarrow$}}#1{}}

\def\Tr{{\rm Tr}}
\def\3#1{{}^3\!#1}\def\4#1{{}^4\!#1}\def\+#1{{}^+\!#1}\def\-#1{{}^-\!#1}
\def\*#1{{}^*\!#1}

\begin{document}

\begintitle{June 1992}{154}{\hfil DEGENERATE EXTENSIONS\hfil\break
\centerline{OF}\hfil\break
GENERAL RELATIVITY}{Ted Jacobson\footnote{jacobson@umdhep.umd.edu} and Joseph
D. Romano\footnote{romano@umdhep.umd.edu} }
General relativity has previously been extended to incorporate degenerate
metrics using Ashtekar's hamiltonian formulation of the theory.  In this
letter, we show that a natural alternative choice for the form of the
hamiltonian constraints leads to a theory which agrees with GR for
non-degenerate metrics, but differs in the degenerate sector from
Ashtekar's original degenerate extension.  The Poisson bracket algebra
of the alternative constraints closes in the non-degenerate
sector, with structure functions that involve the {\it inverse} of the
spatial triad. Thus, the algebra does {\it not} close in the degenerate
sector. We find that it must be supplemented by an infinite number of
secondary constraints, which are shown to be first class (although their
explicit form is not worked out in detail).  All of the constraints taken
together are implied by, but do not imply, Ashtekar's original form of
constraints. Thus, the alternative constraints give rise to a different
degenerate extension of GR. In the corresponding quantum theory,
the single loop
and intersecting loop holonomy states found in the connection representation
satisfy {\it all} of the constraints. These states are therefore exact (formal)
solutions to this alternative degenerate extension of quantum gravity, even
though they are {\it not} solutions to the usual vector constraint.

\vspace{.7cm}


\endtitle

In Ashtekar's hamiltonian formulation of general relativity \cite{Ashtekar1},
the phase space variables are an $SL(2,C)$ spin-connection $A_i{}^{AB}$ and
a density-weighted spatial triad $\otw\s^i{}_{AB}$.\footnote{Lower case latin
indices $i,j,\cdots$ \ are spatial coordinate indices,
while upper case latin indices $A,B,\cdots$ \ are two-component spinor
indices.  The spinor indices are raised and lowered with the anti-symmetric
spinor, $\e_{AB}$, together with its inverse, $\e^{AB}$, according to the
conventions $\l^A:=\e^{AB}\l_B$, $\u_B:=\u^A\e_{AB}$, and $\e_{AB}\e^{AB}:=2$.
An over(under)-tilde is used to indicate a density weight $+(-)1$. We will
also sometimes use a matrix notation suppressing spinor indices. Implicit
contractions are always taken from upper left to lower right, except for the
trace operation, in which the implicit contraction is from lower left to
upper right. For example, $[M,N]_{AB}:=M_A{}^C N_{CB}-N_A{}^C M_{CB}$,
whereas $\Tr\, M:=M_A{}^A=M_{AB}\e^{AB}$.} The constraints of the theory
are
$$\eqalignnotwo{\cG&:=\cD_i\otw\s^i=0,&(1a)\cr
\cV_i&:=\Tr(\otw\s^j F_{ij})=0,\quad{\rm and}&(1b)\cr
\cS&:=\Tr(\otw\s^i\otw\s^j F_{ij})=0,&(1c)\cr}$$
where $\cD_i$ is the covariant spatial derivative operator defined by $A_i$,
and $F_{ij}=2\pd_{[i}A_{j]}+[A_i,A_j]$ is its associated curvature tensor.
The constraints $\cG$, $\cV_i$, $\cS$ are called the Gauss, vector, and scalar
constraints, and the Hamiltonian is simply a linear
combination of these constraints. Note that the vector constraint is {\it not}
bilinear and hence does not generate pure spatial diffeomorphisms. It contains
a term quadratic in $A_i$ which generates a field-dependent, local $SL(2,C)$
transformation. The net operation generated by Poisson bracket with the vector
constraint is a gauge-covariant generalization of the Lie derivative
\cite{Ted1}.

The vector and scalar constraints ($1b,c$) can also be written as a {\it
combined} constraint \cite{Ashtekar2}
$$\cC:=\otw\s^i\otw\s^j F_{ij}=0,\eqno(2)$$
where the two spinor indices have been suppressed.  In terms of $\cS$
and $\cV_i$, \ $\cC$ is given by
$$\cC_{AB}={1\over 2}\e_{AB}\ \cS+\otw\s^i{}_{AB}\ \cV_i.\eqno(3)$$
The vector and scalar constraints imply the combined constraint; conversely,
the anti-symmetric part of the combined constraint
implies the scalar constraint, while its symmetric part implies the vector
constraint only if $\otw\s^i$ is non-degenerate.  Explicitly,
$\cS=\Tr\ \cC$ and $\cV_i=-\Tr(\utw\s{}_i\cC)$, where $\utw\s{}_i$ is the
inverse of $\otw\s^i$ defined by $\otw\s^j{}_{AB}\utw\s{}_i{}^{AB}=\d_i^j$.

Ashtekar's formulation of general relativity (GR) allows one to extend
the standard theory to
include {\it degenerate} triads, since the constraints are polynomial in the
canonical coordinates $(A_i,\otw\s^i)$. Note, however, that the two forms of
the diffeomorphism constraints (($1b,c$) and (2)) seem to define different
extended theories, due to the inequivalence mentioned above.
Put differently, the combined constraint generates diffeomorphisms in
field-dependent directions of the form $N^{AB}\otw\s^i{}_{AB}$.
If $\otw\s^i$ is not invertible, then some directions are not obtained in
this manner.  From one point of view, this is unacceptable, because not all
diffeomorphisms have been included.  However, in the world of degenerate
metrics, it is not so clear {\it a priori} what the rules should be.

One inviolable rule that a constrained hamiltonian theory must satisfy is
that the constraints should be preserved in time by the evolution equations.
Since the Hamiltonian for GR is a combination of these constraints, this means
that the Poisson brackets (PB's) of the constraints must vanish modulo the
constraints.  If this condition is not automatically satisfied, then it must
be enforced by secondary constraints. The PB algebra of the constraints in
Ashtekar's original form (1) closes on itself with
structure functions that are polynomial \cite{Ashtekar1}. Thus, the constraints
(1) are suitable for defining an extended theory that includes degenerate
$\otw\sigma^i$.

In this letter we shall find that the PB algebra of the combined constraints
(($1a$) and (2)) closes in the non-degenerate sector, with structure functions
that involve the {\it inverse} of $\otw\s^i$.\footnote{That the PB algebra of
the combined constraints involves  non-polynomial structure functions was first
noticed in \cite{Ashtekar2}.} Thus, the algebra does {\it not} close in the
degenerate sector, so the combined constraints {\it alone} are not suitable
for defining an extended theory.  Rather, they must be supplemented by an
infinite number of secondary constraints, as will be discussed below.
These secondary constraints are first class, and are implied by, but do not
imply, Ashtekar's original form of the vector constraint. Thus, the
combined constraints yield a different degenerate extension of GR.
\vspace{4mm}

To verify these claims, let us first define constraint functionals
$$\eqalignnotwo{\cC(\utw N)&:=\int_\S\Tr(\utw N\otw\s^i\otw\s^j F_{ij})\quad
{\rm and}&(4a)\cr
\cG(\L)&:=\int_\S\Tr(\L\cD_i\otw\s^i),&(4b)\cr}$$
where $\utw N{}_{AB}$ and $\L_{AB}$ are smearing fields on the
spatial manifold $\S$.
Since the Gauss constraint functional $\cG(\Lambda)$ generates $SL(2,C)$
gauge transformations \cite{Ashtekar1}, we immediately have
$\{\cG(\L),\cG(\L')\}=\cG([\L,\L'])$ and $\{\cG(\L),\cC(\utw N)\}=
\cC([\L,\utw N])$. Thus, the PB of $\cG(\L)$ with any other constraint
functional vanishes modulo that constraint.

Let  $\approx$ denote equality on the constraint surface defined by
($1a$) and (2).  An expression $\approx 0$ is said to {\it weakly} vanish.
To evaluate the PB of two combined constraints, it is
convenient to write $\cC$ as $\textstyle{1\over 2}\cS+\otw\s^i\ \cV_i$.
The  bracket $\{\cC(\utw N),\cV_i\}$ weakly vanishes, because $\cV_i$
generates spatial diffeomorphisms and gauge transformations. The
bracket $\{\cS(\utw N),\cS\}$ weakly vanishes, because it is proportional to
$\otw\s^i\ \cV_i$. Thus, we find
$$\eqalignnotwo{&\{\cC(\utw N),\cS\}\approx -2\cD_j(\cV_i\utw N^{AB})
(\otw\s^{[i}\otw\s^{j]})_{AB}\quad{\rm and}&(5a)\cr
&\{\cC(\utw N),\otw\s^i{}_{AB}\ \cV_i\}\approx 2\cD_j(\utw N\otw\s^{[i}
\otw\s^{j]})_{(AB)}\ \cV_i.&(5b)\cr}$$
Since the RHS of (5) is not guaranteed to weakly vanish when $\otw\s^i$ is
degenerate, we see that the PB algebra of the combined constraints does {\it
not} close in the degenerate sector.  In the non-degenerate sector, the
algebra does close,
but it involves structure functions that are non-polynomial in $\otw\s^i$.
That is, one can write $\cV_i=\d_i^j\ \cV_j = \utw\s{}_i{}^{CD}
(\otw\s^j{}_{CD}\ \cV_j)$, absorbing the inverse $\utw\s{}_i$ as part of the
structure function.\footnote{Although it is not needed for this letter, we
have found that, in the non-degenerate sector, the PB of $\cC(\utw N)$ and
$\cC(\utw M)$ is given explicitly (up to a boundary term that vanishes for
spatially compact or asymptotically flat $\S$'s) by
$\{\cC(\utw N),\cC(\utw M)\}=\cC(\utw K)+\cG(\L)$, where
$$\eqalignnotwo{\utw K{}_{AB}&=-2\left[\utw N{}_{AD}\cD_i
\Big(\utw M^C{}_E(\otw\s^{[i}\otw\s^{j]}){}^{DE}\Big)-\utw M\leftrightarrow
\utw N\right]\utw\s{}_{jBC}\quad{\rm and}&\cr
\L_{AB}&={1\over 2}\utw M{}_C{}^E\utw N{}_{DE} F_{ij}{}^{CD}
(\otw\s^i\otw\s^j)_{AB}.&\cr}$$
(Note that the structure function associated with $\utw K_{AB}$ involves the
inverse $\utw\s{}_j$.) One might have hoped that the structure of the PB
algebra of the combined constraints would have had a simple geometric
interpretation, providing insight into the dynamics of GR.
Unfortunately, the form of $\{\cC(\utw N),\cC(\utw M)\}$ and the expressions
for $\utw K{}_{AB}$ and $\L_{AB}$ do not suggest (to us) any geometric
interpretation.}

Since the Hamiltonian is just a linear combination of $\cC$ and $\cG$,
the conditions that the PB's in (5) vanish are secondary constraints,
required to hold if the combined constraint $\cC=0$ is to be preserved under
time evolution. These secondary constraints may be imposed as additional
conditions on the initial data, or they may be implemented as restrictions
on the Lagrange multiplier $\utw N$ that enters the Hamiltonian.
To stay as close as possible to standard GR, we will decline to restrict the
Lagrange multiplier;\footnote{Integrating by parts in ($5a$), and expanding
the derivative in ($5b$) using the product rule, one sees that the weak
vanishing of the secondary constraints (5) is equivalent to the condition
$\utw N^{AB}\phi_{BC}\approx 0$, where $\phi_{BC}:=\cD_j(\otw\s^{[i}
\otw\s^{j]})_{BC}\ \cV_i$.  In the generic case, when $\phi$ is
invertible, this implies $\utw N^{AB}\approx 0$, so there can be no evolution
at all. If $\phi$ is degenerate, then $\phi_{BC}=\l_B\l_C$, and $\utw N^{AB}$
must weakly be of the form $\u^A\l^B$ for some $\u^A$.  Equivalently,
$\utw N^{AB}\approx\r^{AC}\phi_C{}^B$ for some $\r^{AC}$.}
instead we will require that the secondary constraints weakly vanish for all
$\utw N$.

The vanishing of ($5a$) for all $\utw N$ is weakly equivalent to the
vanishing of ($5b$) for all $\utw N$.  Thus, we can take the secondary
constraints to be given by
$$\phi^{(1)}_{AB}(\utw N):=\{\cC(\utw N),\otw\s^i{}_{AB}\}\ \cV_i=
2\cD_j(\utw N\otw\s^{[i}\otw\s^{j]})_{(AB)}\ \cV_i=0,\eqno(6)$$
for all $\utw N$.
Since $\phi=0$ must also be preserved in time, we must further require, as
additional constraints, the vanishing of $\{\cC,\phi\}$,
$\{\cC,\{\cC,\phi\}\}$, {\it etc.} (Note that we can effectively ignore the
Gauss constraint functionals in this analysis, since the PB of $\cG(\L)$ with
any constraint vanishes modulo that constraint.) This process seems to continue
forever, apparently yielding an infinite number of secondary constraints.
Using $\{\cC,\cV_i\}\approx 0$, the secondary constraints take the form
$$\eqalignnotwo{\phi^{(1)}_{AB}(\utw N)&:=\{\cC(\utw N),\otw\s^i{}_{AB}\}\
\cV_i
=0,&(7a)\cr
\phi^{(2)}_{AB}(\utw N)&:=\{\cC(\utw N),\{\cC(\utw N),\otw\s^i{}_{AB}\}\}\
\cV_i
=0,&(7b)\cr
&{\it etc.}&\cr}$$
We will now show that this chain of secondary constraints is first class, and
is weaker than Ashtekar's original form of the vector constraint.

To verify that the algebra of all the constraints is first class, it suffices
to show that constraints of the form $\cC$, $\{\cC,\cC\}$,
$\{\cC,\{\cC,\cC\}\}$, $\{\cC,\{\cC,\{\cC,\cC\}\}\}$ {\it etc.} have PB's that
vanish modulo these constraints. (As before, we can ignore the Gauss constraint
functionals in this analysis.) This follows from the Jacobi identity, by an
induction argument:
Let $I_n$ denote a nested bracket of $n$ $\cC$'s. If, for all $n$,
$\{I_n,I_k\}$ is a linear combination of constraints of the form $I_{n+k}$,
then
$\{I_n,I_{1+k}\}=\{I_n,\{I_1,I_k\}\}=\{I_1,\{I_n,I_k\}\}+\{I_k,\{I_1,I_n\}\}
=I_{1+n+k}-\{I_{1+n},I_k\}$.

The phase space for the degenerate extension of GR that adopts the constraints
in the form (1) originally given by Ashtekar is clearly contained within that
defined by ($1a$), the combined constraint (2), and the secondary constraints
(7). To see that the latter theory is {\it inequivalent} to the former, it
suffices to note that every constraint other than the Gauss constraint has at
least one factor like $\otw\s^{[i}\otw\s^{j]}$ in it.  Any $\otw\s^i$ of the
form $\otw\s^i{}_{AB}=\otw V{}^i S_{AB}$ will satisfy (2) and all the secondary
constraints, and some $\otw\s^i$ of this form will also satisfy the Gauss
constraint but not the vector constraint. Thus, the phase space for the latter
theory is evidently {\it larger}.  Exactly how much larger is not clear,
however, as we have not worked out the proper counting of degrees of freedom.

It is a rather unfamiliar and possibly unhealthy feature of this alternative
degenerate extension of GR, that there are an infinite number of constraints
per space point. Indeed, the vanishing of all of the constraints seems only
to ensure the preservation of the constraints in a Taylor series expansion
in time. If the initial data is not analytic, it seems that the constraints
are {\it not} necessarily preserved by the evolution.  If this is so, either
non-analytic data would have to be excluded, or additional constraints would
have to be required.

This new degenerate extension of GR can be quantized, leading to an
inequivalent quantum gravity theory.  An illustration of this inequivalence
will now be discussed, using Dirac quantization in the connection
representation. Note that
since the constraints are first class, they generate gauge
transformations, and there is no need to reduce the phase space and pass to
Dirac brackets.  Thus, $\otw\s^i$ is still conjugate to $A_i$, and in the
connection representation the operator $\widehat{\otw\s}{}^i$
is still represented by $\d/\d A_i$.

Now a number of quantum states $\Psi$ have been found that satisfy the Gauss
and scalar constraints.  These states
are functionals of the connection, given by the trace of the holonomy
around a fixed closed loop in space, and generalizations thereof that involve
intersecting loops \cite{TedLee,Viqar,BerndJorge}. They do {\it not}
satisfy the vector constraint, since the fixed loops are not diffeomorphism
invariant. On the other hand, these loop states satisfy all four of the
constraint equations in the combined form $\widehat{\cC}\Psi=0$, as well as
all of the secondary constraints. This is because these states are annihilated
by the operator corresponding to $\otw\s^{[i}\otw\s^{j]}$, which appears in
all of these constraints. If the combined constraint is ordered with this
operator on the right, and if all of the secondary constraints are given by
operator commutators of the $\cC$'s, then all of the constraint operators
will annihilate these states.
Thus, for such operator orderings, the loop states are exact (formal)
solutions to this alternative degenerate extension of quantum gravity.

One way to understand the discrepancy between the two
forms of the constraints (($1b,c$) or (2)) acting on the loop states is to
note, as mentioned earlier,
that the combined constraint (2) generates diffeomorphisms only in
directions of the form $\Tr(\utw N\otw\s^i)$. The triad is degenerate in the
loop states, since $\widehat{\otw\s}{}^i(x)\Psi$ is proportional to the loop
tangent vector at $x$.\footnote{That $\widehat{\otw\s}{}^i(x)\Psi$ vanishes
for all $x$
not on the loop, and is therefore even more degenerate, is not relevant to
the point being made here. The state satisfies the vector constraint
as well at such points.} For a  non-intersecting loop state there is only a
single tangent vector at each point. The non-intersecting loop state is
indeed invariant under diffeomorphisms that are {\it tangent} to the loop,
so it satisfies the combined form of the constraints. For the intersecting
states, there is more than one tangent vector at the point of intersection,
so the situation is more complicated. The diffeomorphisms under which
the state is invariant are a quantum superposition of diffeomorphisms
along the intersecting curves.

Finally, it is of course possible to define degenerate extensions of GR by
covariant, 4-dimensional action principles.  For instance, one can use one
of the polynomial, first-order actions that have been used to derive Ashtekar's
hamiltonian formulation \cite{Sam,Ted2,CDJM}. When the metric is degenerate,
the canonical reduction of these actions does not go through as usual. We do
not
know what the relation is between the degenerate extensions of hamiltonian
GR discussed here and such covariant lagrangian extensions.

\vspace{1cm}

\noindent ACKNOWLEDGEMENTS

\vspace{.3cm}

We would like to thank Abhay Ashtekar for initially proposing this
problem, and for suggestions that improved the presentation of this paper.
We would also like to thank Marc Henneaux for helpful discussions.
This work was supported in part by NSF grant PHY91-12240.

\end{document}